\documentstyle[preprint,aps,epsf]{revtex}

\begin{document}
\draft
\preprint{KANAZAWA 97-03,  
\ April, 1997}  
\title{Inter-meson Potentials in Dual Ginzburg-Landau Theory
}
\author{
Hiroaki Kodama$^{a}$
\footnote[2]{E-mail address : {\tt h-kodama, matsubara, suzuki
@hep.s.kanazawa-u.ac.jp}}, 
Yoshimi Matsubara$^{b}$  $^{\dag}$, 
Shigeyoshi Ohno$^{c}$ 
\footnote[3]{E-mail address : {\tt ohno@uitec.ac.jp}} 
and Tsuneo Suzuki$^{a}$  $^{\dag}$,
}
\address{$^{a}$
Department of Physics, Kanazawa University, Kanazawa 920-11, Japan
}
\address{$^{b}$
Nanao Junior College, Nanao, Ishikawa 926, Japan
}
\address{$^{c}$
Shiga Polytechnic College, Shiga 523, Japan
}
\maketitle

\begin{abstract}
We calculate inter-meson potentials numerically by solving classical 
equations of motion derived from 
Dual Ginzburg-Landau (DGL) Theory.
Inter-meson potentials in DGL theory are shown to be 
similar to those of the string-flip model and well reproduce behaviors
of the short-range interaction at the classical level.
We also compare our results with those from lattice 
QCD Monte carlo calculations.

\end{abstract}

\newpage

\section{Introduction}

Quark confinement is one of the main ploblems that should be solved 
in hadronic physics. The linear potential 
between quark and anti-quark at long distance 
can be 
calculated by Monte Carlo calculations of lattice QCD and 
the non-relativistic 
quark potential 
model explain the low-lying hadron spectrum with the infinitely
rising potentials.\cite{q.p1}

Quark potentials in multi-hadron system should 
also be investigated. 
A difficulty 
appears when one try to understand multi-hadron systems 
in the framework of linear rising quark potential model. 
It gives a long-range attraction, called color van der Waals force, 
between two color-singlet hadrons which seems to contradict experimental 
data of nucleon-nucleon scattering.\cite{van-der} 

In treating multi-hadron system, the 
string flip model was proposed phenomenologically.\cite{st.flip} \ 
In the string flip model, hadrons do not interact each other 
except when the strings flip into another combination. 
Hence the string flip model is by construction 
free from color van der Waals force. 
But it is not so straightforward to understand the string flip model
starting from QCD.

Dual Ginzburg-Landau (DGL) theory is an infrared effective theory 
of quark confinement derived from QCD.\cite{DGL} \  
'tHooft and Mandelstam conjectured that QCD vacuum may be dual to 
superconductor and that monopoles play the analogous role of Cooper pairs 
in QCD.\cite{'tHooft1,Mandels} 
'tHooft proposed abelian projection to extract relevant 
dynamical variable at low energy in QCD and suggested that 
the variable is color magnetic monopole.\cite{'tHooft2}
The abelian projection is a prescription of partial gauge fixing which 
reduces $SU(3)$
gauge group to $U(1)\times U(1)$ and monopoles appear as a point like 
singularity. Following Bardacki and Samuel we can integrate out 
monopole trajectories.\cite{Bar-Sam} \  Introducing dual vector gauge 
fields,\cite{Zwan} \  
we can get Dual Ginzburg-Landau Lagrangian that is an abelian effective 
theory with magnetic monopoles.

Meson, baryon and spin-dependent 
baryon potentials have been 
calculated numerically in DGL theory.\cite{numerical-pot,numerical-pot2} \ 
The model also explains the characteristic features of 
finite-temperature transition of pure QCD found by lattice calculations.
\cite{DGL2} \  
Monopole condensation and chiral symmetry breaking 
is also discussed.\cite{DGL3}  

In this paper, we treat 
multi-hadron system (for simplicity meson-meson system) in DGL 
theory. We numerically solve classical equations of motion derived 
from DGL theory and obtain static potentials  \(V_2\) in two body 
systems and \(V_4\) in four body systems. Inter-meson potential is 
evaluated from \(E_4=V_4-2V_2\). We also compare our results 
from DGL theory with those from lattice Monte Carlo calculations done 
by Green et al..\cite{Green}

\section{inter-meson potential in DGL theory}

DGL theory is represented by a Lagrangian
\begin{eqnarray}
   {\cal L}_{DGL} & = & -\frac 14 {(\vec{H}_{\mu\nu})}^2+\sum_{\alpha=1}^3
   \left\{ {\left| ( \partial_\mu + ig{\vec{\epsilon}}_\alpha\cdot
   {\vec{C}}_\mu ) \chi_\alpha \right| }^2 -\lambda 
   {\left( {| \chi_\alpha |}^2 -v^2 \right)}^2\right\}
   \label{eqn:1s1}
\end{eqnarray}
with the constraint \(\sum_{\alpha =1}^3 \mbox{arg}(\chi_\alpha)=0 \) , 
where \( \vec{C}_\mu=(C_\mu^3,C_\nu^8) \) is 
the dual abelian vector field with respect to
$U(1)$\(\times\)$U(1)$ and \(\chi_\alpha (\alpha=1,2,3)\) 
are the complex scalar monopole fields
 which couple to \( \vec{C}_\mu \) covariantly under magnetic 
$U(1)$\(\times\)$U(1)$.\cite{Bar-Sam} \  
Here \(\vec{\epsilon}_\alpha\) are the root 
vectors : \( \vec{\epsilon}_1=(1,0),\ 
\vec{\epsilon}_2=(-1/2,-\sqrt 3/2),\ 
\vec{\epsilon}_3=(-1/2,\sqrt 3/2)\).   
\(\vec{H}_{\mu\nu}\) is defined as
\begin{eqnarray}
   {\vec H}_{\mu\nu} & = & \partial_\mu{\vec C}_\nu
   -\partial_\nu{\vec C}_\mu+\epsilon_{\mu\nu\lambda\sigma}\int d^4y
   \;n^\lambda ( n\cdot\partial )^{-1} (x-y){\vec j_{\mbox {ex}}}^\sigma (y),
   \label{eqn:1s2}
\end{eqnarray}
where \(n^\lambda\) is an arbitrary 
constant four vector and 
\(\vec{j}^\sigma_{\mbox {ex}}=(j_3^\sigma,j_8^\sigma)\) is 
an external color-electric
current. We have already assumed monopole condensation and abelian 
dominance\cite{DGL} \ which is supported by 
Monte Carlo simulations of QCD.\cite{ab.domi} \  
It is noted that this form is just dual to relativistic Ginzburg-Landau 
theory with an external magnetic current.

We consider for simplicity the following symmetric configurations 
of four static quarks in three dimensional space :
\begin{eqnarray}
   {\vec{j}}_{\mbox{ex}}^{\mu} (r)
    =  \vec{Q}_2~g^{\mu 0} & \biggl\{ &
       \delta \left( x \right)\delta \left( y -\frac D2 \right)
       \delta \left( z -\frac R2 \right)\nonumber\\
    &-&\delta \left( x \right)\delta \left( y -\frac D2 \right)
       \delta \left( z +\frac R2 \right)\nonumber\\
    &-&\delta \left( x \right)\delta \left( y +\frac D2 \right)
       \delta \left( z -\frac R2 \right)\nonumber\\
    &+&\delta \left( x \right)\delta \left( y +\frac D2 \right)
       \delta \left( z +\frac R2 \right)\biggr\},
   \label{eqn:1s3}
\end{eqnarray}
where
\begin{eqnarray}
   \vec{Q}_2&=&(Q^3_2,Q^8_2)
   =\left( -\frac{e}{2} \ ,-\frac{e}{2\sqrt 3}\right) .
   \label{eqn:1s3.5}
\end{eqnarray} 
Here color-electric coupling \(e\) 
is related to \(g\) by Dirac quantization condition 
\(eg=4 \pi \).

We now look for static solutions in which the time components are neglected. 
Adopting a unitary gauge \({\mbox{arg}}(\chi_\alpha)=0\) ,
 we get equations of motion : 
\begin{equation}
   \mbox{\boldmath $\nabla$}\times\vec{\mbox{\boldmath $E$}}
   -2g^2\sum_{\alpha =1}^{3} {\vec{\varepsilon}}_{\alpha}
   (
    {\vec{\varepsilon }}_{\alpha}\cdot\vec{\mbox{\boldmath $C$}} )
   {| \chi_\alpha |}^2=0,
   \label{eqn:1s4}
\end{equation}
\begin{equation}
   \Delta {| \chi_\alpha |}
   -2g^2
   ({\vec{\epsilon }}_{\alpha}\cdot\vec{\mbox{\boldmath $C$}} )
   {| \chi_\alpha |}
    -2\lambda({| \chi_\alpha |}^2-v^2){| \chi_\alpha |}=0,
    \qquad(\alpha=1,2,3).
   \label{eqn:1s5}
\end{equation}
The energy density \(\cal{E}\) is written as
\begin{equation}
   {\cal{E}}=\frac 12 {\vec{\mbox{\boldmath $E$}}}^2
   +\sum_{\alpha =1}^{3} \left\{ {\left( \nabla | {\chi}_\alpha | \right)}^2
   +g^2{\left( 
   {\vec{\varepsilon}}_{\alpha}
   \cdot\vec{\mbox{\boldmath $C$}} \right)}^2 {|{\chi}_{\alpha}|}^2
   +\lambda {\left( {|{\chi}_{\alpha}|}^2-v^2 \right)}^2 \right\}.
   \label{eqn:1s6}
\end{equation}
Here the color-electric field is defined as
\begin{eqnarray}
   \vec{\mbox{\boldmath $E$}}=\mbox{\boldmath $\nabla$}\times
  \vec{{\mbox{\boldmath $C$}}}+\vec{{\mbox{\boldmath $E$}}}_S,
   \label{eqn:1s7}
\end{eqnarray}
where \(\vec{{\mbox{\boldmath $E$}}}_S\) is the singular string field which vanishes everywhere except on 
the Dirac strings. It comes from the last term of Eq.~(\ref{eqn:1s2}).

If we require 
\(\vec{\mbox{\boldmath $E$}}\to 0\)     
and \(\chi_\alpha\to v\) at the spatial infinity,
our gauge choice makes the 
$\vec{\mbox{\boldmath $C$}}$ field vanish at this region 
as seen from Eq.~(\ref{eqn:1s4}). This means that the string singularity does 
not extend to infinity but connects the quark charges 
in the unitary gauge. There are various ways to connect the charges.

Here we consider the Dirac quantization condition which guarantees 
unobservability of the string. We can change the string 
location by a single-valued gauge transformation, if the integral 
of the dual vector field     along an infinitesimal closed 
loop is quantized as follows :  
\begin{equation}
   \lim\oint \vec{\mbox{\boldmath $C$}}
   d\mbox{\boldmath $l$}=\sum_{i=1}^{3}{\xi}_i{\vec{Q}}_i\ ,
   \label{eqn:1s8}
\end{equation}
where \(\xi\)     is an integer and \(eg=4\pi\)    \label{eqn:1s1}
is used.
\hspace{-4mm}
\footnote[1]{Consider an infinitesimal closed loop around a 
Dirac string. After a gauge transformation around the closed loop,
the dual vector gauge fields
$\vec{{\mbox{\boldmath $C$}}}'=\vec{{\mbox{\boldmath $C$}}}+
\mbox{\boldmath $\nabla$}\vec{\bf\Lambda}_m$
become regular : 
$\lim\oint (\vec{\mbox{\boldmath $C$}}+\mbox{\boldmath $\nabla$}
\vec{\mbox{\boldmath $\bf\Lambda$}}_m)
d\mbox{\boldmath $l$}=0$.
Also the $\chi_\alpha$ fields remain single-valued :
$g\vec{\epsilon}_\alpha\cdot(\oint \mbox{\boldmath $\nabla$}
\vec{\mbox{\boldmath $\bf\Lambda$}}_m
d\mbox{\boldmath $l$})=2\pi n$.
Hence we see
$g\vec{\epsilon}_\alpha\cdot(\lim\oint \vec{\mbox{\boldmath $C$}}
d\mbox{\boldmath $l$})=2\pi n$
which is the generalized Dirac quantization condition 
$\vec{m}\cdot\vec{q}=2\pi n$.
Using Eq.~(\ref{eqn:1s3.5}), we get Eq.~(\ref{eqn:1s8}).
}
Condition (\ref{eqn:1s8}) selects 
two-type configurations of the Dirac string (Figs.~1\ (a)\ \mbox{and }(b))
 out of those which 
connect the charges (Figs.~1\ (a),\ (b)\ \mbox{and } (c)). 
Note that in the unitary gauge the 
location of the color-electric flux coincides with that of the 
Dirac string where the value of the monopole fields \(\chi_\alpha\) 
vanishes.

From \(j_3^\mu =\sqrt{3}j_8^\mu\) it is 
expected that there exists a solution satisfying 
\(C^3_\mu =\sqrt 3C^8_\mu\equiv C_\mu \).
Then two equations of (\ref{eqn:1s5}) \((\alpha=1,\ 2)\) become identical, 
so that we get the same solution for \(\chi_1\) and \(\chi_2\)
\(( | \chi_1 |=|\chi_2|\equiv\chi) \) . Moreover 
we can solve the remaining equation of (\ref{eqn:1s5}) \((\alpha=3)\) 
with the condition 
that the static energy is minimized :
\begin{equation}
   |\chi_3|=v.
   \label{eqn:1s9}
\end{equation}
To solve Eqs.~(\ref{eqn:1s4}) and (\ref{eqn:1s5}) 
numerically, we have to avoid infinite 
quantities such as the string singularities. Following 
Ball and Caticha,\cite{Ball} \  
we separate
\begin{equation}
   {\mbox{\boldmath $C$}}={\mbox{\boldmath $C$}}_D+
   {\tilde{ \mbox{\boldmath $C$} }}.
   \label{eqn:1s10}
\end{equation}
Here \({\mbox{\boldmath $C$}}_D\) represents 
the Coulomb field generated by the four charges. In the case of (a) type 
string location in Fig.~1, it is written as
\begin{eqnarray}
   {\mbox{\boldmath $C$}}_D(\vec{r}) & = & 
   \frac{Q^3_2}{4\pi}\frac{\cos{\theta}_1+1}
   {|{\mbox{\boldmath $r$}}-{\mbox{\boldmath $r_1$}}|
   \sin{\theta}_1}{\hat{\phi}}_1
   -\frac{Q^3_2}{4\pi}\frac{\cos{\theta}_2+1}
   {|\mbox{\boldmath $r$}-\mbox{\boldmath $r_2$}|
   \sin{\theta}_2}{\hat{\phi}}_2\nonumber\\
   &&-\frac{Q^3_2}{4\pi}\frac{\cos{\theta}_3+1}
   {|\mbox{\boldmath $r$}-\mbox{\boldmath $r_3$}|
   \sin{\theta}_3}{\hat{\phi}}_3
   +\frac{Q^3_2}{4\pi}\frac{\cos{\theta}_4+1}
   {|\mbox{\boldmath $r$}-\mbox{\boldmath $r_4$}|
   \sin{\theta}_4}{\hat{\phi}}_4,
   \label{eqn:1s11}
\end{eqnarray}
where the notation is explained in Fig.~2~. The new fields 
\({\tilde{ \mbox{\boldmath $C$} }}\) and
\(\nabla\times {\tilde{ \mbox{\boldmath $C$} }}\)     
are well behaved everywhere.
The field equations are reduced to
\begin{eqnarray}
   &&{\nabla}^2 \chi - g^2 ({\mbox{\boldmath $C$}}_D 
   + {\tilde{\mbox{\boldmath $C$}}})^2
   \chi = 2\lambda ({\chi}^2 - 1)\chi,
   \label{eqn:1s12}\\
   &&{\nabla}^2 \tilde{\mbox{\boldmath $C$}} 
   - {\nabla} \cdot ({\nabla}{\tilde{\mbox{\boldmath $C$}}} )=
   3g^2 {\chi}^2 ({\mbox{\boldmath $C$}}_{D} 
   + {\tilde{\mbox{\boldmath $C$}}}).   \label{eqn:1s13}
\end{eqnarray}
Eq.~(\ref{eqn:1s10}) not only eliminates the singularities from the equations 
for \({\tilde{\mbox{\boldmath $C$}}}\)    
but also enables a separation of the infinite Coulomb self-energy. 
The static potential energy in four body system without the 
self-energy is given by
\begin{eqnarray}
   V_4(R,D)=&&{\frac{e^2}{6\pi}} \left\{ 
   - \frac 1R - \frac 1D 
   +\frac {1}{\sqrt{(R^2+D^2)}} \right\} \nonumber\\
   &&+ \int d^3x
   \left\{ -2g^2{\chi}^2 \mbox{\boldmath $C$}\cdot
   \tilde{\mbox{\boldmath $C$}} + 2\lambda(v^4-{\chi}^4) \right\}.
   \label{eqn:1s14}   
\end{eqnarray}
As the boundary condition at the space-like infinity, we have required 
\(\mbox{\boldmath $C$}\to 0\)     
and \(\chi\to v\)      . 
This means the vacuum of the dual superconductor is realized 
at the infinite boundary.

The static potential $V_2(R)$ in two body system can be evaluated in a similar 
way. Finally we get inter-meson potential $E_4$ using the equation
\begin{equation}
   E_4(R,D) = V_4(R,D)-\left\{ 
     \begin{array}{rl}
       2V_2~(R), & \quad \mbox{if } R<D,\\
       2V_2~(D), & \quad \mbox{if } R>D.
     \end{array}\right.
   \label{eqn:1s15}
\end{equation}

\section{Numerical Method}

As we treat the system numerically, we rescale the variables as $ x'\equiv vx, \mbox{ }
\tilde{\mbox{\boldmath ${C}$}}'\equiv v^{-1}\tilde{\mbox{\boldmath ${C}$}}, 
\mbox{ }\mbox{\boldmath $C$}_D'\equiv v^{-1}
\mbox{\boldmath $C$}_D,\mbox{ }\chi '\equiv v^{-1}\chi\ $ and
$V'(R,D) \equiv v^{-1} V(R,D)$      . 

The spatial boundary conditions are 
imposed on the surface of sufficiently large but finite volume. Since the 
system has (anti)symmetries under \(x\to -x\),
\(y\to -y\) or \(z\to -z\)     , we can restrict 
ourselves to the region ( \(x\geq 0,\ y\geq 0,\ z\geq 0\)  ) 
with the appropriate continuity 
conditions. The region is discretized into an \(N_x\times N_y\times N_z\) 
lattice. 
Since a fine lattice is not needed near the boundary surface, 
we have used lattices 
whose lattice spacing becomes larger as the distance 
from the quark sources. We have used the Gauss-Seidel method applying the successive 
overrelaxation technique. Furthermore, we have adopted the following 
procedure to get rapid convergence. We work first on a coarse lattice and 
next employ the solution as the intial configuration on a finer lattice 
which is constructed by putting new sites between the original lattice 
sites. And we repeat this process. The largest values of \(N_x,N_y\), and 
\(N_z\) 
attained are 80, 120 and 120, respectively.

The procedure of changing the lattice size shows us the 
lattice spacing dependence of the energy \(V_4\) and \(V_2\). 
It is nicely fitted by \(V_{4,2}=c_0+c_1a^2+c_2a^4\) where 
\(a\) is a typical 
lattice spacing. This dependence gives the continuum limit of the energy.

Varying the inter-quark distances \(R\) and \(D\), 
we get the potential \(V_4\) and \(V_2\).

\section{Results and discussion}

We first illustrate the results of the inter-quark potential.
As shown in Fig.~3, inter-quark potential in DGL theory 
is Yukawa-like in the short distance and linear in the long distance.
DGL theory realizes quark confinement.

To compare our results based on DGL theory with those from $SU(2)$ lattice QCD 
calculation obtained by Green et al.,\cite{Green} \ we study 
the $SU(2)$ case. In $SU(2)$ case, Eqs.~(\ref{eqn:1s12}) and 
(\ref{eqn:1s13}) are almost 
the same as in $SU(3)$ case except that the coupling  
$3g^2$ is replaced by $2g^2$ in Eq.~(\ref{eqn:1s13}).
The four quark potential is modified into
\begin{eqnarray}
   V(R,D)=&&{\frac{e^2}{2\pi}} \left\{ 
   - \frac 1R - \frac 1D 
   +\frac {1}{\sqrt{(R^2+D^2)}} \right\} \nonumber\\
   &&+ \int d^3x
   \left\{ -g^2{\chi}^2 \mbox{\boldmath $C$}\cdot
   \tilde{\mbox{\boldmath $C$}} + \lambda(v^4-{\chi}^4) \right\}
   \label{eqn:1s14-3}   
\end{eqnarray}
Our results in $SU(3)$ case is qualitatively similar to those in $SU(2)$ 
case. To compare these results, we have to adjust the parameters in both 
results. Free parameters are 
\(e,\lambda,
v\)\enskip in DGL theory 
and \(a_{2.4}\) which is the lattice spacing at \(\beta=2.4\) in 
$SU(2)$ lattice QCD.
Inter-quark potential \(V_2(R)\) calculated 
numerically in DGL theory gives the scaled string tension 
\(\sigma/v^2=6.2809(2)\), 
whereas lattice $SU(2)$ QCD Monte Carlo 
simulation gives the scaled string tension 
\(\sigma a^2_{2.4}=0.0699\).\cite{Green} \  
The string tension \(\sigma =(440\ \mbox{MeV})^2 \) 
determines \(v\) and \(a_{2.4}\) to be 175.567(3)
MeV and 0.1179 fm, respectively. 
Because all charged gluons as well as neutral 
gluons contribute equally in the short distance, 
the Coulomb term of \(V_2(R) \) in DGL theory\cite{numerical-pot} \  
should be equal to one third of 
the term \(-0.249/R\) derived from the lattice data 
.
Hence we set \(e=2.04\). In Ref.~\cite{border}, it is found that 
the QCD vacuum is near the border between 
type 1 and 2. Hence we take 
\(\lambda=\frac 12(\frac{4\pi}e)^2\) in accordance with 
the Ginzburg-Landau parameter \(\kappa=\sqrt{\lambda}/g=1/\sqrt{2}\).

Now we can calculate the inter-meson potential. We consider 
both cases of Dirac string location (Fig.~1 (a) and (b)).
Using Eq.~(\ref{eqn:1s15}) we evaluate the inter-meson potential 
\(E_4(R,D)\) in both cases and choose the lower one. 
In Fig.~4  the results are shown when \(R\) is fixed. Two mesons 
hardly interact each other in the region \(D\gg R\).
But as two mesons are getting closer, interaction between 
them becomes sizable at some value of 
\(D\). The maximally interacting point is just the square (\(R=D\)) point.
Still decreasing \(D\) makes Dirac strings flip to take the 
holizontal direction 
and the interaction gets weaker.
This behaviour is similar to the ansatz made in the string 
flip model.\cite{st.flip}

To see whether these inter-meson potentials in DGL theory are long-range 
or short-range, \(E_4(R,D)\) in the region \(D \geq R\) 
is fitted by the function 
\begin{equation}
  -\alpha\frac{e^{-\delta D}}{D^{\gamma}}
   \label{eqn:4s17}
\end{equation}
when \(R=0.1179\) fm.
This function reproduces well the data when 
\(\delta=1.43(1)\)\ GeV and\ \(\gamma=1.145(1)\). So the inter-meson 
potential in DGL theory is short-range.

When \(R\) is taken larger, the peak value of interaction becomes 
smaller (Fig.~5). The case \(R=D\) is shown in Fig.~6.
In Fig.~7 we also show the potential in the case that 
Dirac strings run in the same direction. 
In this case the inter-meson potential is repulsive as expected.

We also compare our results with those from lattice QCD 
.\cite{Green}  Both results 
are similar qualitatively
as seen from Figs.~4, 5 and 6. The degree of falling at 
\(R\approx D\) in DGL theory is a little weaker than that in lattice QCD. 
This difference may come from the assumption that we 
have neglected the off-diagonal gluons whose contribution is  
sizable in the short distance.

In Fig.~8  the color electric flux distribution is shown.
The color electric flux spreads in all 
direction in Coulomb phase $(v=0)$, 
while it is squeezed 
between quarks which belong to the same meson 
in the confinement phase $(v\ne 0)$.
This illustrates the interactions between mesons are short-range.
Thus DGL theory reproduces both properties of quark 
confinement and asymptotic separability of 
mesons at the classical level. The potentials in DGL theory 
is linear in the color electric 
flux direction, whereas Yukawa-like short-range in the direction 
perpendicular to the flux.

Recently an attempt to determine the couplings 
in the abelian effective monopole action 
has been performed. Monopole action can be 
derived numerically from the monopole current configurations given after 
abelian projection on the lattice.\cite{shiba} \  
This action is equivalent to non-compact abelian Higgs 
action on the dual lattice in the long-range region.\cite{Smit} \  
It is very interesting to 
fix the couplings in DGL theory directly from QCD.

\vspace{1cm}
This calculation were performed on Fujitsu VPP500 
at the institute of Physical and Chemical Research (RIKEN).

\input epsf

\begin{figure}
\epsfxsize=\textwidth
\begin{center}
\leavevmode
\epsfbox{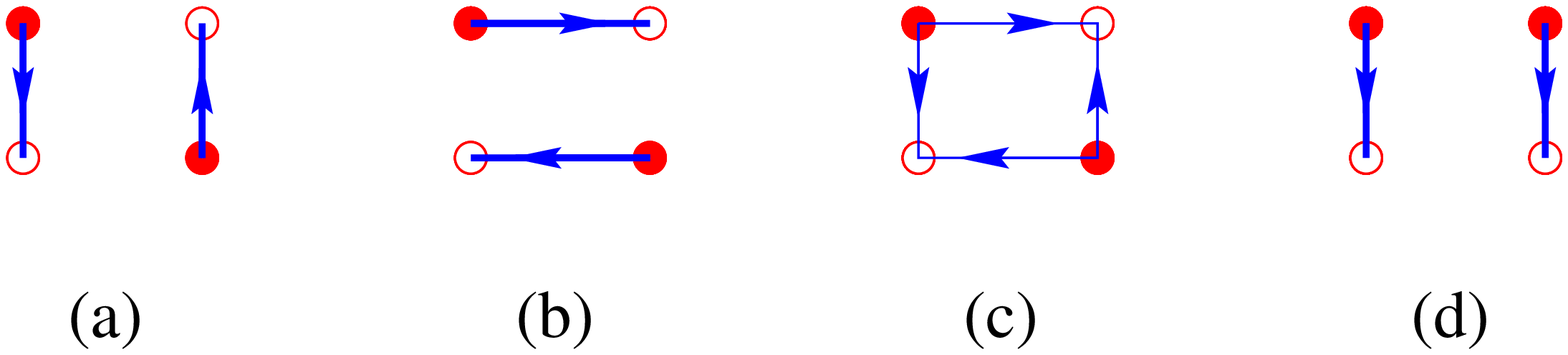}
\end{center}
\caption{
Locations of Dirac strings. (a), (b) each charge has 
one Dirac string. Two Dirac strings run in the opposite 
direction. (c) each charge has two Dirac strings. (d) each charge 
has one Dirac string. Two Dirac strings run in the same direction.
}
\label{lloop}
\end{figure}

\begin{figure}
\epsfxsize=\textwidth
\begin{center}
\leavevmode
\epsfbox{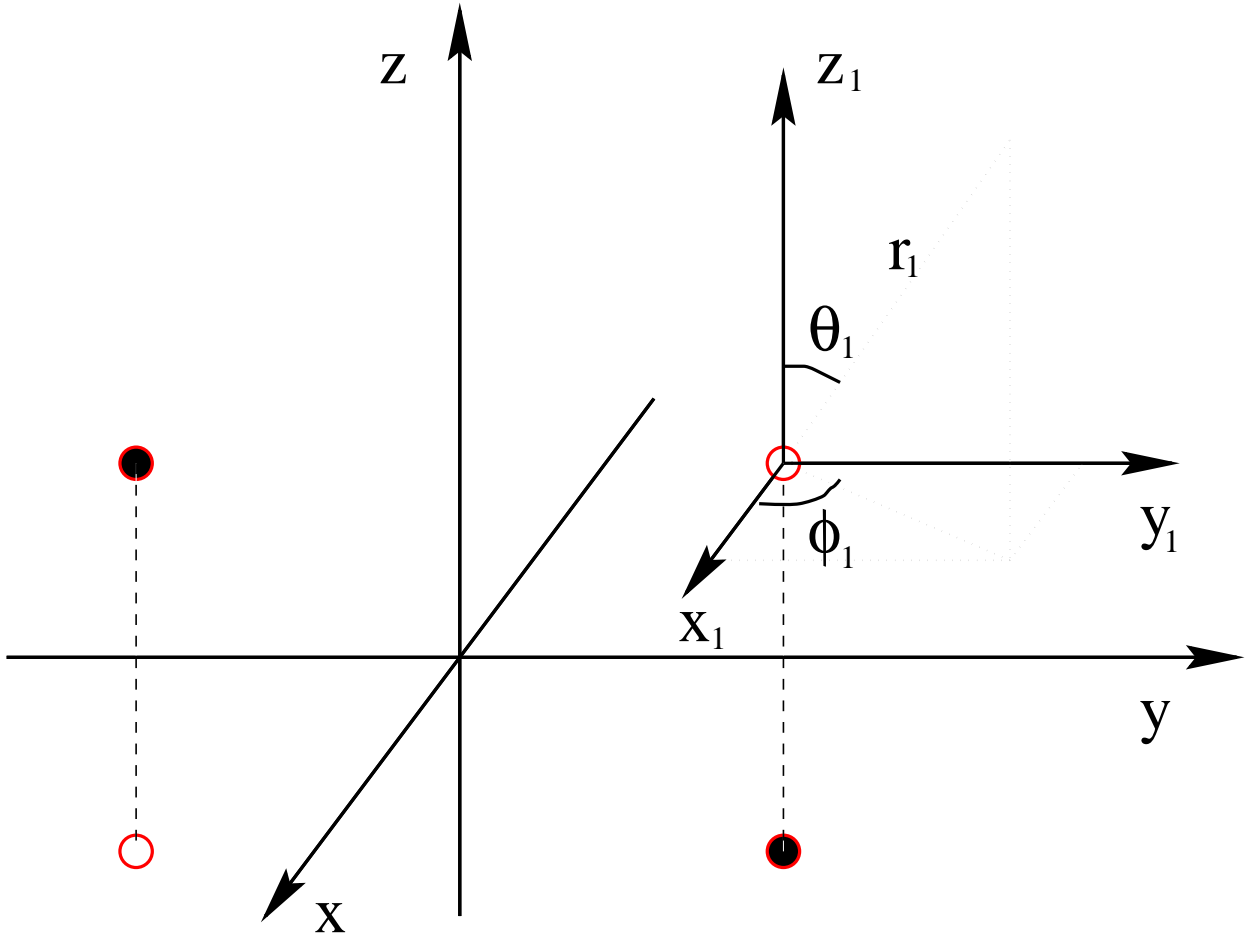}
\end{center}
\caption{
Coordinates used in Eq.~(\protect\ref{eqn:1s11}).
}
\label{lloop}
\end{figure}

\begin{figure}
\epsfxsize=\textwidth
\begin{center}
\leavevmode
\epsfbox{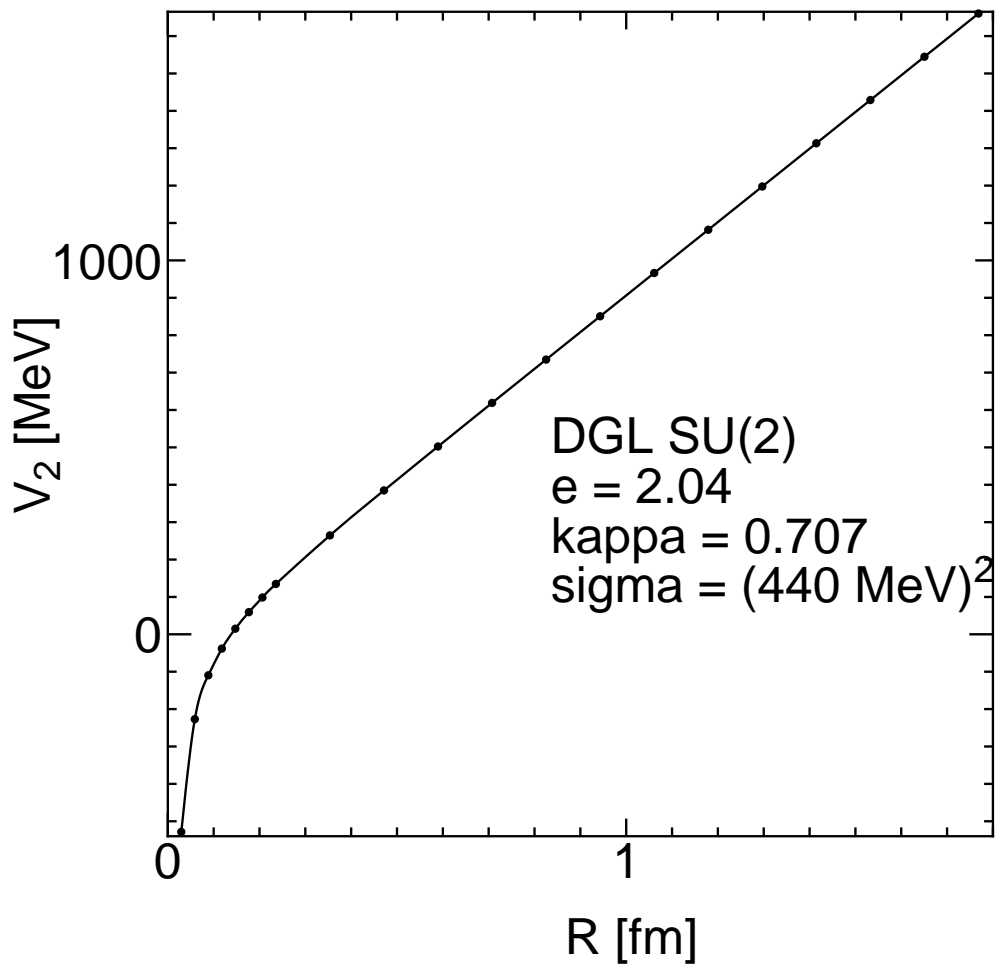}
\end{center}
\caption{
Inter-quark potential derived from $SU(2)$ DGL theory when 
$e$=2.04, $\kappa=0.707$,\ 
$\sigma={(440 \  \mbox{MeV})}^2 $. 
}
\label{lloop}
\end{figure}

\begin{figure}
\epsfxsize=\textwidth
\begin{center}
\leavevmode
\epsfbox{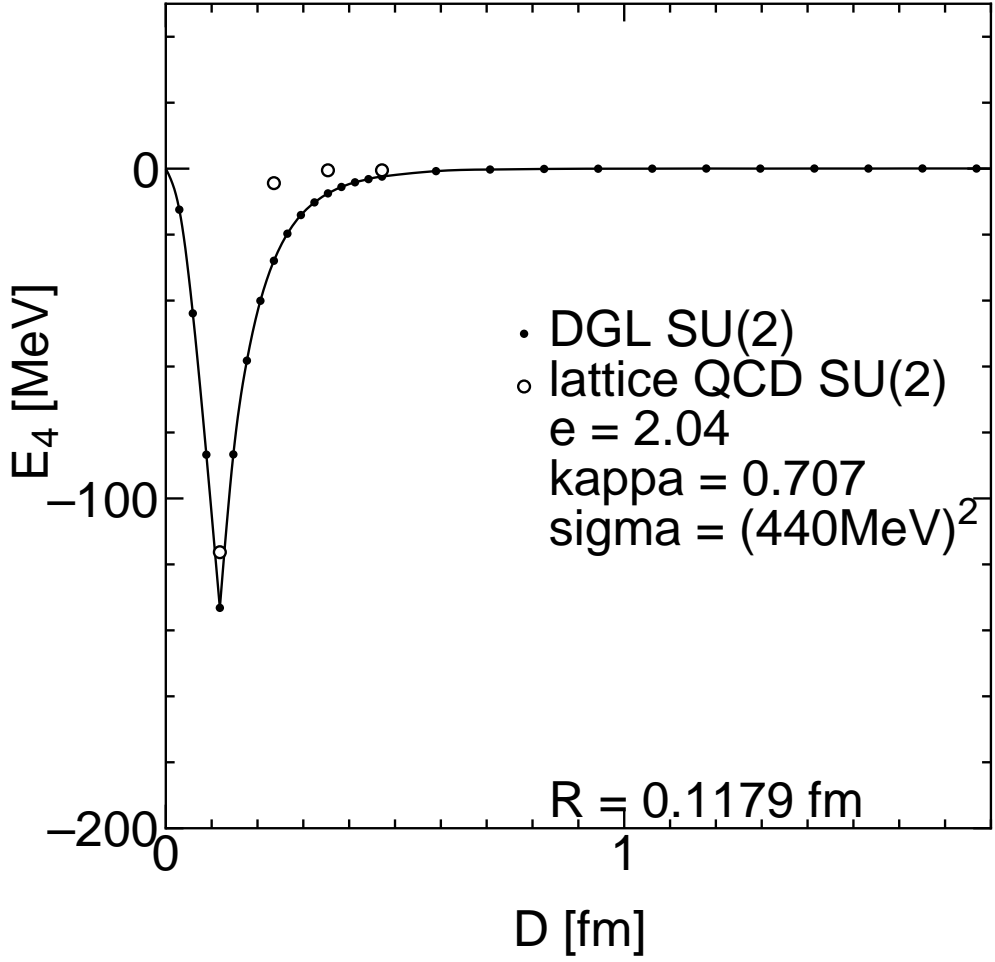}
\end{center}
\caption{Inter-meson potential dreived from $SU(2)$ DGL theory (dot) when 
$e$=2.04, $\kappa$=0.707,\ 
$\sigma={(440 \  \mbox{MeV})}^2$. 
The open circle denotes potential 
from $SU(2)$ lattice QCD in Ref.~\protect\cite{Green}. 
Quark charges are located at the corner of 
the rectangle and  
two Dirac strings run in the opposite direction (Figs.~1 (a) and (b)).
$R$ is fixed to be $0.1179\ \mbox{fm}$.
}
\label{asst}
\end{figure}

\begin{figure}
\epsfxsize=\textwidth
\begin{center}
\leavevmode
\epsfbox{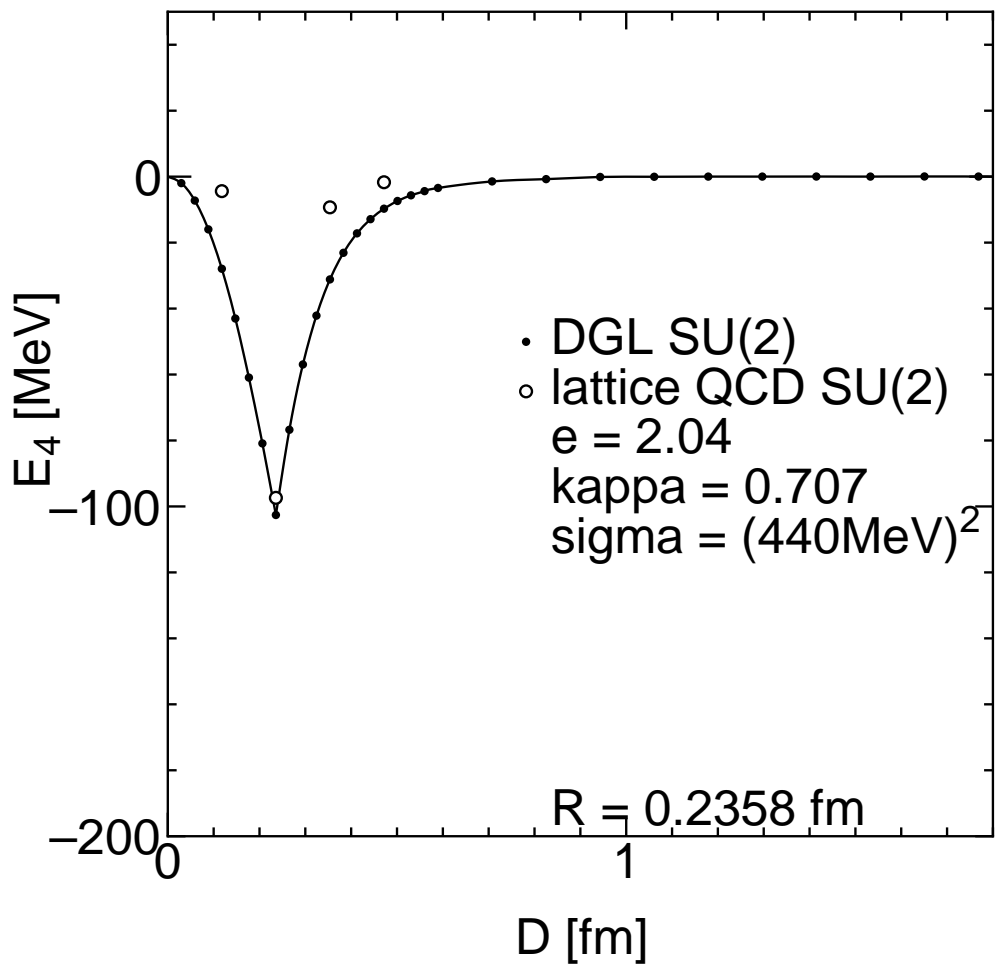}
\end{center}
\caption{Inter-meson potential dreived from $SU(2)$ DGL theory 
when $R$ is fixed to be $0.2358\ \mbox{fm}$.
}

\label{msst}
\end{figure}

\begin{figure}
\epsfxsize=\textwidth
\begin{center}
\leavevmode
\epsfbox{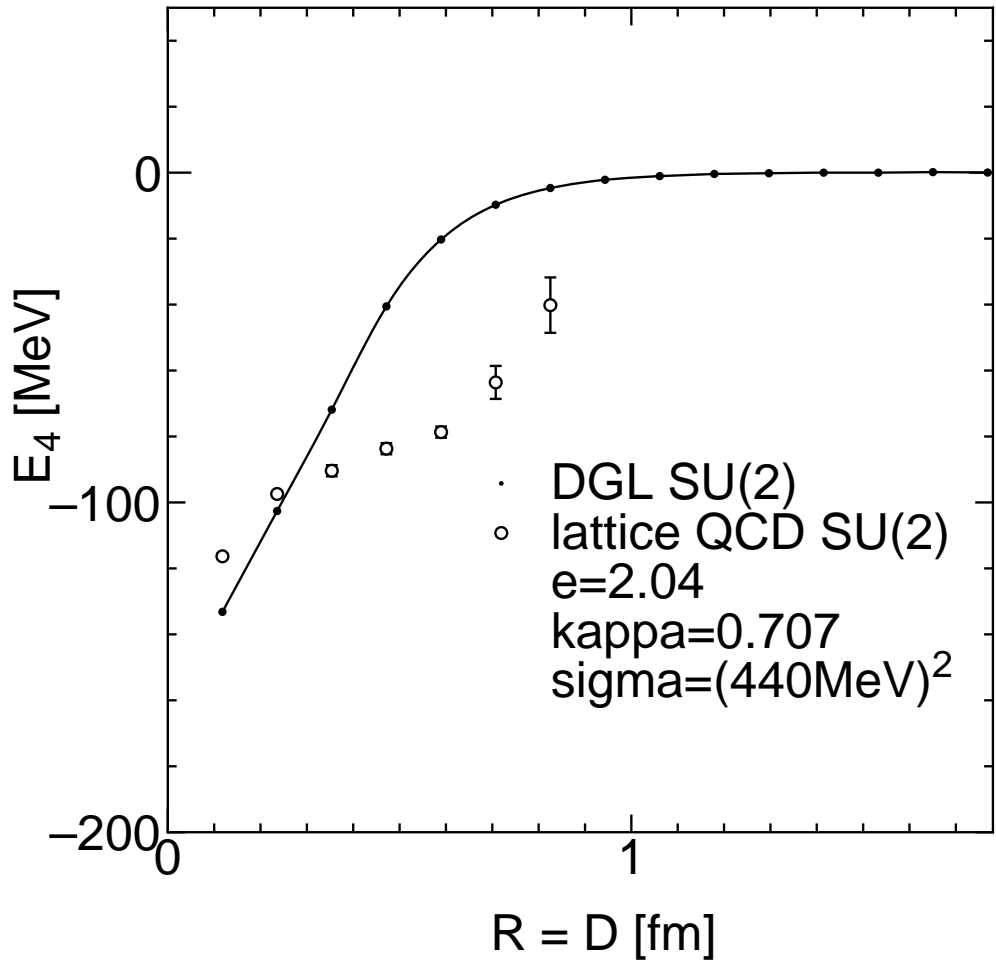}
\end{center}
\caption{Inter-meson potential dreived from $SU(2)$ DGL theory 
in the $R=D$ case.
}
\label{mgsc}
\end{figure}

\begin{figure}
\epsfxsize=\textwidth
\begin{center}
\leavevmode
\epsfbox{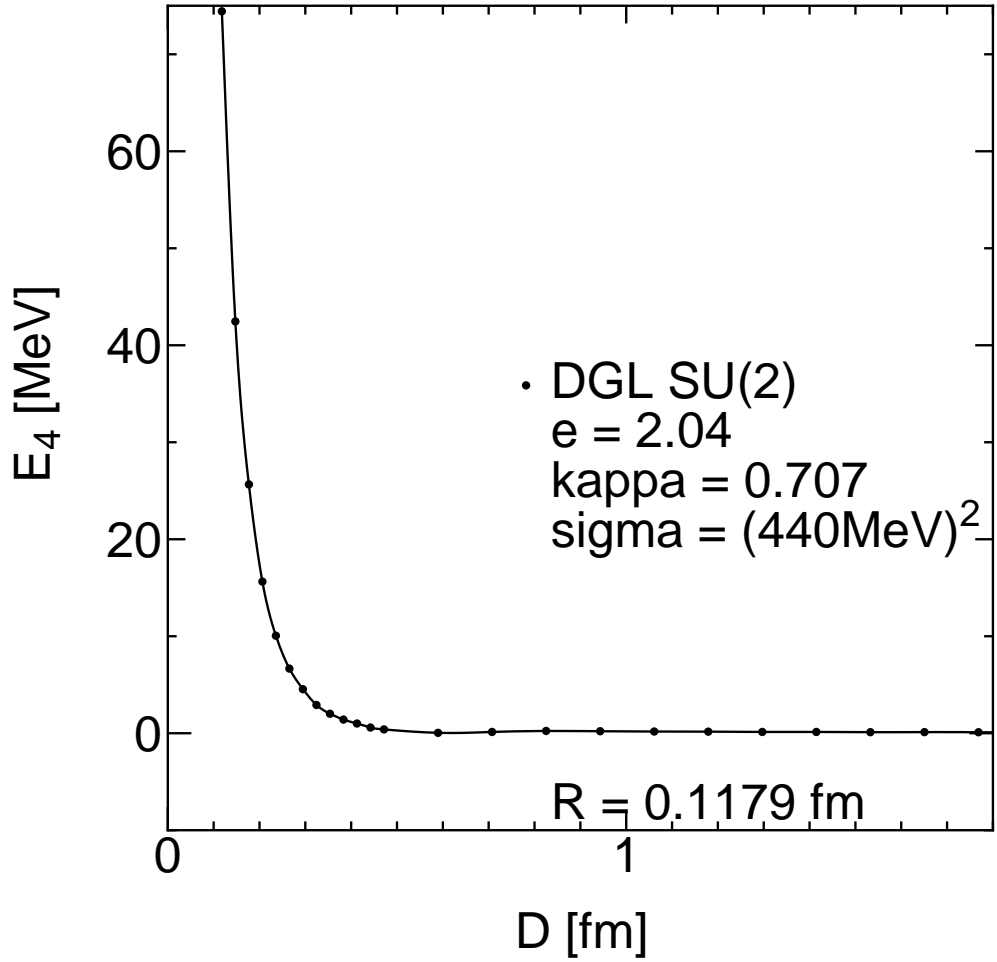}
\end{center}
\caption{Inter-meson potential dreived from $SU(2)$ DGL theory
when two Dirac srtings run in the same direction (Fig.~1 (d)).
$R$ is fixed to be $0.1179\ \mbox{fm}$.
}
\label{msst}
\end{figure}

\begin{figure}
\begin{center}
\epsfysize=13cm
\leavevmode
\epsfbox{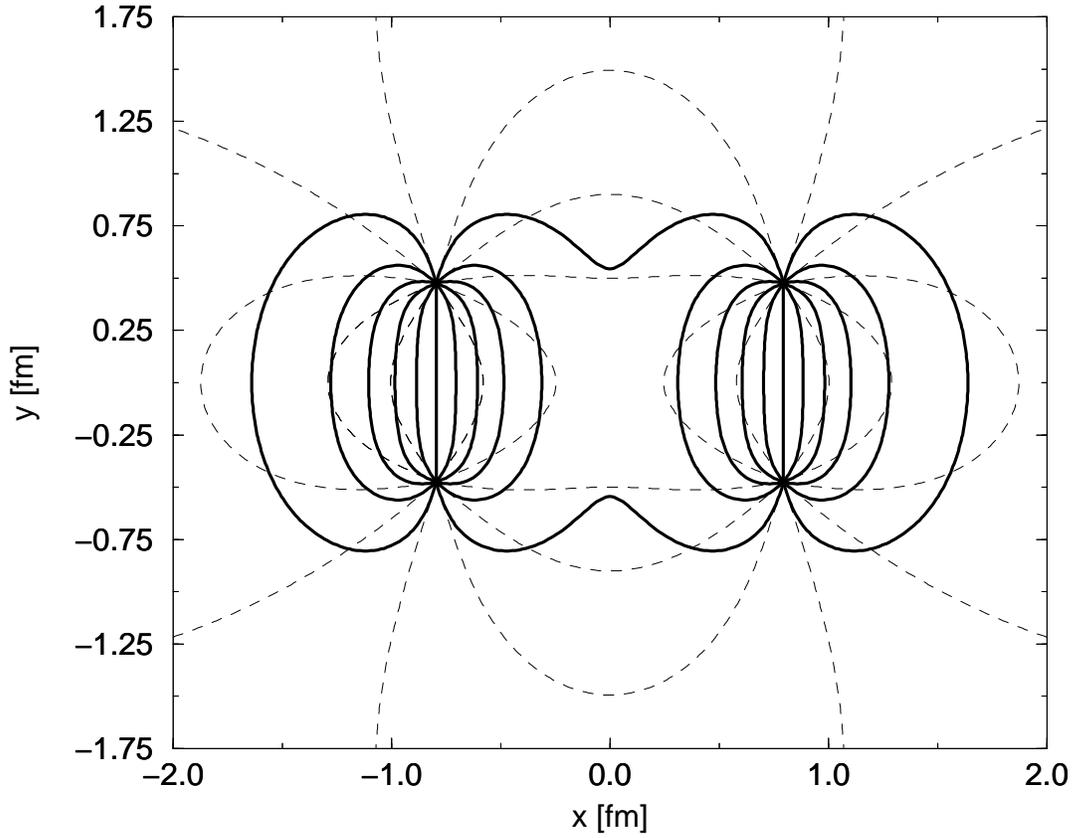}
\end{center}
\vspace{2cm}
\caption{
Color electric flux in confinement phase $v\neq 0$ (solid line) 
and that in Coulomb phase $v=0$ (dashed line).
Color electric charges are located at
$(x,y)$=($\pm$0.795 fm, $\pm$ 0.445 fm). 
Parameters are taken to be 
$e$=2.04, $\kappa$=0.707,\ 
$\sigma={(440 \  \mbox{MeV})}^2 $ .
}
\label{mgsc}
\end{figure}

\end{document}